\documentstyle[preprint,aps,pra]{revtex}
\begin{document} 
\title {THE AB-INITIO SIMULATION OF THE LIQUID Ga-Se SYSTEM}
\author{J.  M.  Holender and M.  J.  Gillan }
\address{Physics Department, Keele University
\\ Keele, Staffordshire ST5 5BG, U.K.}
\date{\today}
\maketitle
\begin{abstract}
Ab-initio dynamical simulation is used to study the liquid
Ga-Se system at the three concentrations Ga$_2$Se,
GaSe and Ga$_2$Se$_3$ at the temperature 1300~K.
The simulations are based on the density functional pseudopotential
technique, with the system maintained  on the Born-Oppenheimer surface
by conjugate gradients minimization. We present results for the
partial structure  factors and radial distribution functions, which
reveal how the liquid structure depends on the composition.
Our calculations of the electrical conductivity $\sigma$ using the
Kubo-Greenwood approximation show that $\sigma$ depends very strongly
on the composition.
We show how this variation
of $\sigma$ is related to the calculated electronic density of states.
Comparisons with recent experimental determinations of the structure
and conductivity are also presented.
\end{abstract}

\section{Introduction}
The properties of many binary liquids depend very strongly on
the composition (e.g.\cite{review}).
We have chosen the metal-selenium systems to study these phenomena.
There are experimental measurements showing the very strong and
unusual changes in the  electronic properties of these systems
\cite{review,agse,inse}.
We  use ab-initio molecular dynamics (AIMD) to study the liquid structure
and the relation between this structure and the electronic properties.
In this paper we report results for the liquid Ga-Se system
(Ag-Se is discussed in a separate paper \cite{agsemd}).
This mixture of a good metal (gallium) and a poorly conducting
material (selenium) enables us to study the changes in the electronic
properties when we go from a metallic system to an insulating one.
In order to have a complete picture of the changes in this system we have
also performed simulations on the pure
liquid elements Ga \cite{gal} and Se \cite{se}.
The ab-initio approach allows us to study simultaneously
the structure and the
electronic properties, and our calculations are free from
any assumptions about the form of the interatomic interaction
(for other AIMD work on liquids, see e.g.
refs.\cite{sti89,zha90,wij94,kre94,sch95}).
There are no experimental data for the liquid Ga-Se system published so far,
but we shall present a comparison with
preliminary data from the Bristol group \cite{gase}.

\section{Technique}
Our ab-initio molecular dynamics
method is based on the Car-Parrinello approach \cite{cp}, but instead of
using `fake dynamics' we bring the system to the Born-Oppenheimer surface at
every ionic configuration using
conjugate gradients minimisation.
We deal with the semi-metallic character of  our system by treating the
occupation numbers as additional variables. The details of our method
are described elsewhere \cite{agsemd,gal}.
Once the electronic ground states is reached, the quantum forces
are calculated; we can then apply the standard algorithm of classical
molecular dynamics.

The ab-initio simulations are very computationally demanding and,
even when using
supercomputers, there are very strong limitations on
the size of the system and the length of the simulations.
We use 60 atoms in a cubic box with the periodic boundary conditions.
We have performed calculations for three concentrations, namely
Ga$_2$Se (40 Ga atoms and 20 Se atoms), GaSe (30 Ga atoms and 30  Se
atoms) and Ga$_2$Se$_3$ (24 Ga atoms and 36 Se atoms).
The calculations were done at densities interpolated between pure Ga
and pure Se (such a simple interpolation at room temperature
predicts densities of solid GaSe and Ga$_2$Se$_3$ within a few percent).

We use non-local pseudopotentials and a
plane-wave cut-off of 150~eV, with only the gamma point used to sample
the  Brillouin zone.
All simulations were done at the
temperature 1300\,K.
The integration time-step for the ion dynamics was 3~fs, and we equilibrated
our system (at  each concentration) for over 1~ps and the ensemble
averages were calculated from the next 4~ps.
\section{Results}
\subsection{Structural properties}
In Figure 1 we compare our neutron weighted structure factors with
preliminary experimental results \cite{gase}.
Our structure factors were calculated directly in $k$-space so the
resolution for small $k$ is very low and the noise is substantial.
The agreement is reasonably good: the positions of the
peaks are correct and the general dependence of peak heights
on concentration is also reproduced correctly. There
is some disagreement in peak heights, especially for
Ga$_2$Se, but one should bear in mind
the small size of our system.

It is very useful to analyse changes in the partial structure factors
and the radial
distribution functions even though we do not have the experimental
results to compare with.
In figure 2 we show the partial Faber-Ziman structure factors together
with the total structure factor for all three
concentrations in the region of the first
peaks. One can see that the structure of the partial $S(k)$ is much
richer than of the total one but there is a high degree of
cancellation of the peaks (the partial structure factors contribute to
the total one with the different prefactors). The variation in the
height of the  peak at about 2~\AA$^{-1}$ can be traced to
changes in the degree of
cancellation of the Se-Se and Ga-Se partials at this $k$-vector.

The partial radial distribution functions can be calculated
directly in real space.
In Figure 3 we show the partial radial distribution functions and their
evolution with increasing concentration of selenium.
All of them change markedly. In the Ga-Ga rdf a peak appears at about
2.45~\AA.
In Ga$_2$Se and GaSe, the Se atoms are well separated and the first
peak in the rdf is at just below 4~\AA. For Ga$_2$Se$_3$ we see a clear peak
at about 2.45~\AA\ (there is a very small peak at this position for
GaSe). This distance corresponds to the Se-Se distance in pure
liquid and solid selenium. On average there are four Se-Se bonds in our
system at this concentration.
This number is low because we are at the stoichiometric
concentration. We also observe the growth of the first (and only)
peak in the Ga-Se rdf.

As a useful characterization of our rdf results, we show in Table 1
the average coordination number of different types of
atoms corresponding to a cutoff of 3~\AA.

\subsection{Electronic properties}
The electronic densities of states (together with the results for pure
gallium at 1000~K) are compared in Figure 4.
The changes in the electronic  density of states are substantial.
We start from the almost
free-electron like behaviour for pure gallium and end up with a gap
for Ga$_2$Se$_3$.  This is a real gap in the density of states, and not just
a pseudo-gap or a conduction gap. Interestingly, we found that
we had to run the simulations for a few picoseconds before the gap opened.
During this long equilibration the radial distribution function
remained the same while we saw a steady changes in the energy eigenvalues.
With increasing concentration of selenium, a
large peak builds up between -15~eV and -10~eV.
These are $s$ states of selenium. A detailed analysis of the
electronic density of states, localization of electrons
and the pseudo-charge distribution will be published
elsewhere \cite{gasemd}.

Given the major variations of the density of states,
it is not surprising to find that the electrical conductivity
changes dramatically with
the concentration.
We calculated the d.c. conductivities using the Kubo-Greenwood formula
for the a.c. conductivity
$\sigma(\omega)$ and by extrapolating this to
$\omega=0$. There are many approximations involved so one does not
expect perfect agreement with experiment.
For pure gallium the calculated conductivity is 20~000 $\Omega^{-1}$cm$^{-1}$,
for Ga$_2$Se it is about 1000 $\Omega^{-1}$cm$^{-1}$, for GaSe this value
drops down to about 100 $\Omega^{-1}$cm$^{-1}$ and for Ga$_2$Se$_3$
it is  almost zero. The corresponding experimental values are:
28~000 \cite{garho}, 1000, 25 and 3 $\Omega^{-1}$cm$^{-1}$ \cite{gase}.

\section{Conclusions}
The main conclusions are as follows. The reasonable agreement
of our calculated neutron-weighted structure factors with preliminary
experimental results confirms the reliability of our AIMD simulations.
Our study of the partial structure factors shows that the form of the
neutron-weighted structure factor involves a large degree of
cancellation between features in the partials. The rdfs reveal
that at low Se concentrations, the Se atoms are well separated from
each other, but at the Ga$_2$Se$_3$ composition the
formation of Se-Se short-distance bonds plays a significant role.
The electronic density of states varies strongly with composition,
and increase of Se content causes the opening of a true gap at
the Fermi level.

{}~

The work of JMH is supported by EPSRC grant GR/H67935. The computations
were performed on the Fujitsu VPX240 at Manchester Computer
Centre under EPSRC
grant GR/J69974.
Analysis of the results was performed using distributed hardware
provided under EPSRC grants GR/H31783 and GR/J36266. Discussions with
J. E. Enderby and A. C. Barnes were important in the early stages of
this work, and
we are grateful to S. B. Lague and A. C. Barnes for making their
experimental results available before publication.

\begin{figure}
\caption{The neutron weighted structure factor
$S(k)$ of Ga$_2$Se , GaSe and Ga$_2$Se$_3$.
Solid line and circles connected
by dotted line represent simulation and experimental
values \protect \cite{gase} respectively. Data for Ga$_2$Se are
shifted up 0.5 mb/atom, for Ga$_2$Se$_3$ down by 0.5 mb/atom.}

\label{sf}
\end{figure}

\begin{figure}
\caption{The partial Faber-Ziman structure factors
and the total structure factor for three
concentrations of Ga-Se.}
\end{figure}

\begin{figure}
\caption{The partial radial distribution functions for three
concentrations of Ga-Se.}
\end{figure}

\begin{figure}
\caption{Electronic density of states for pure liquid gallium,
Ga$_2$Se, GaSe and Ga$_2$Se$_3$.}

\label{ds}
\end{figure}
\begin{table}
\caption{The average coordination numbers for Ga-Se alloys and
positions of the first peaks in the radial distribution fuction,
in brackets the interatomic distances and the coordination numbers
of the corresponding solid phase. Distances in \AA.}

\begin{tabular}{cccccccc}
alloy&$r_{Ga-Ga}$&$n_{Ga-Ga}$&$r_{Ga-Se}$&$n_{Ga-Se}$ &$n_{Se-Ga}$&
$r_{Se-Se}$&$n_{Se-Se}$ \\ \hline
Ga$_2$Se & 2.54 & 2.41 & 2.44 & 1.46 & 2.92 & 3.95 & 0.01 \\
GaSe & 2.51 (2.44) & 1.12 (1) & 2.44 (2.45)
& 2.66 (3) & 2.66 (3) & 3.95 (4.75) & 0.02 (0)\\
Ga$_2$Se$_3$ & 2.46 (3.85) & 0.62 (0)& 2.43 (2.35)
& 3.39 (4) & 2.26 (2.67) & 3.95 (3.85) & 0.30 (0)\\
\end{tabular}
\end{table}


\begin {references}

\bibitem{review} J. E. Enderby and A. C. Barnes, Rep. Prog. Phys. {\bf
53} (1990) 85.
\bibitem{agse} S. Ohno, A. C. Barnes and  J. E. Enderby, J. Phys.:
Condens. Matt. {\bf 2} (1990) 7707.
\bibitem{inse} T. Okada and S. Ohno, J. Non-Cryst. Solids
{\bf 156-158} (1993) 748.
\bibitem{agsemd} F. Kirchhoff, J. M. Holender and M. J. Gillan,
Phys. Rev. Lett., submitted.; idem., J. Non-cryst. Solids, this volume
\bibitem{gal} J. M. Holender, M. J. Gillan, M. C. Payne and  A. D.
Simpson, Phys. Rev. {\bf 52} at press.
\bibitem{se} F. Kirchhoff, M. J. Gillan and J. M. Holender,
J. Non-cryst Solids, this volume.
\bibitem{sti89} I. \v{S}tich, R. Car and M.  Parrinello, Phys.  Rev.
Lett. {\bf 63}, (1989) 2240.

\bibitem{zha90} Q. M.  Zhang, G.  Chiarotti, A.  Selloni, R. Car
and M. Parrinello, Phys. Rev. B\ {\bf 42}, (1990) 5071.

\bibitem{wij94} G. A. de Wijs, G. Pastore, A. Selloni and
W. van der Lugt, Europhys. Lett.\ {\bf 27}, (1994) 667.

\bibitem{kre94} G. Kresse and J. Hafner, Phys. Rev.  B\ {\bf 49},
(1994) 14251.

\bibitem{sch95} M. Sch\"one, R. Kaschner and G. Seifert, J. Phys.:
Condens.  Matter\ {\bf 7}, L19 (1995).

\bibitem{gase} S. B. Lague and A. C. Barnes, this volume.
\bibitem{cp} R.  Car and M.  Parrinello, Phys.  Rev.  Lett.\ {\bf
55} (1985) 2471.
\bibitem{gasemd} J. M. Holender and M. J. Gillan,
Phys. Rev. B, to be submitted.
\bibitem{garho} G. Ginter, J. G. Gasser, and R. Kleim, Philos. Mag.
B {\bf54} (1986) 543.

\end{references}
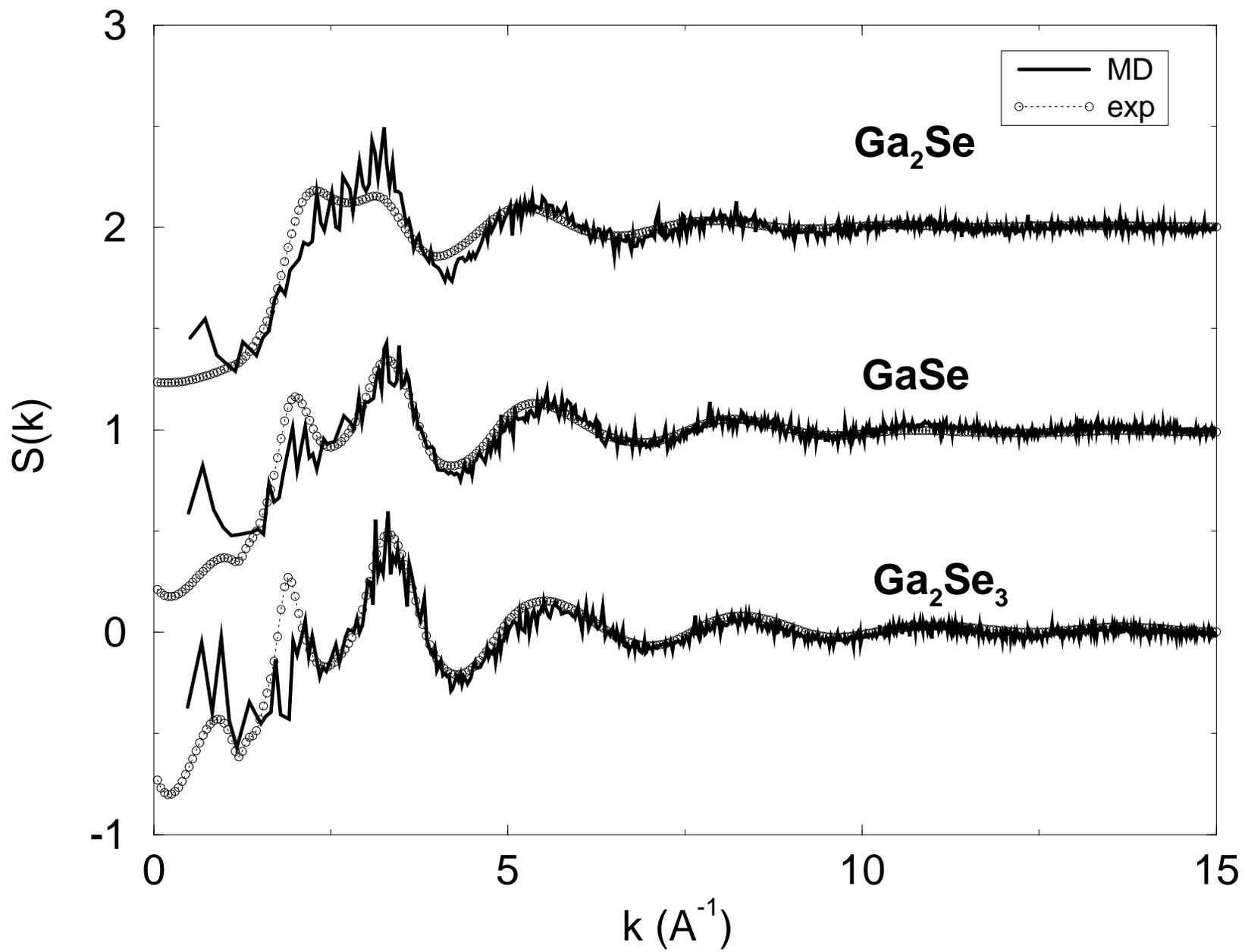
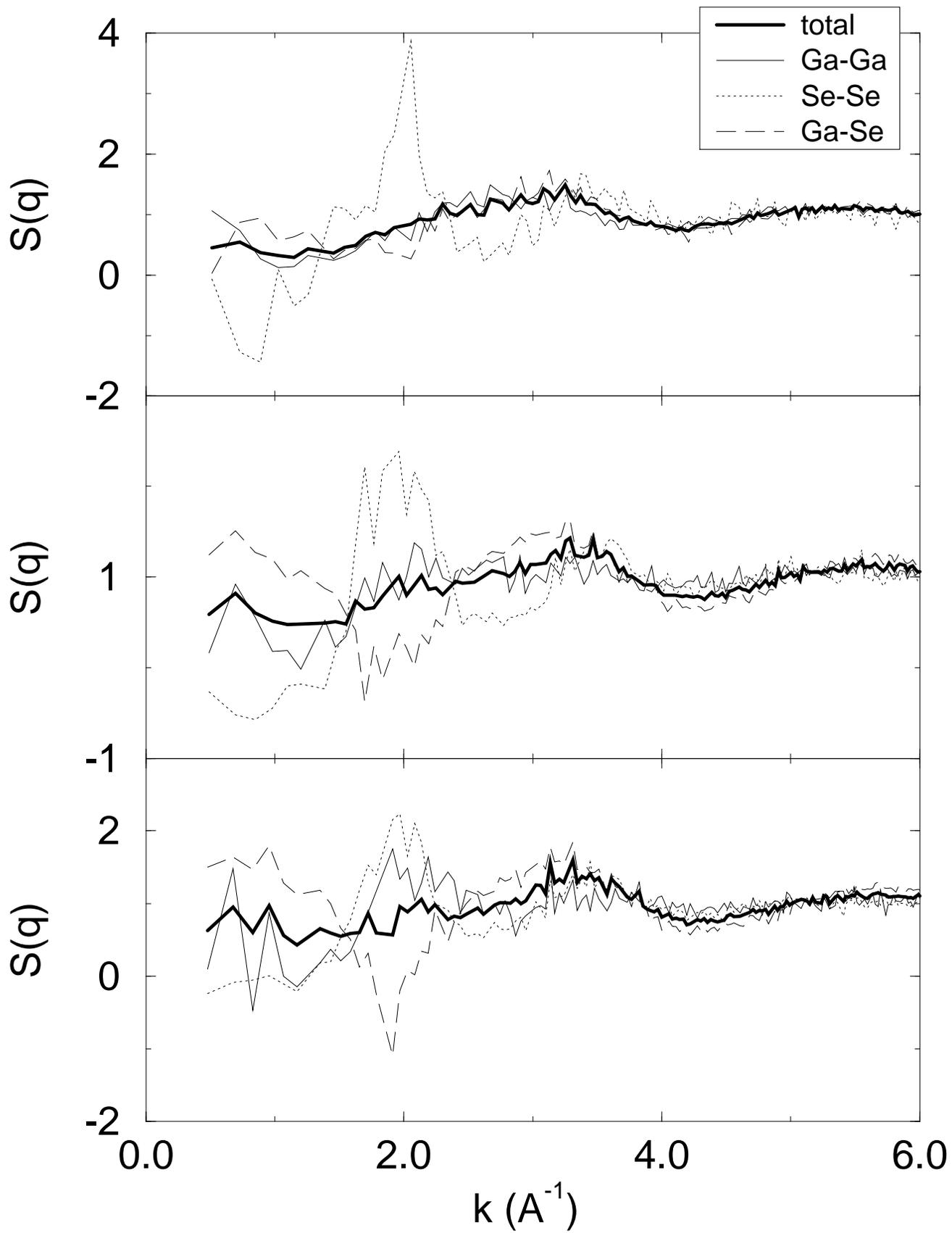
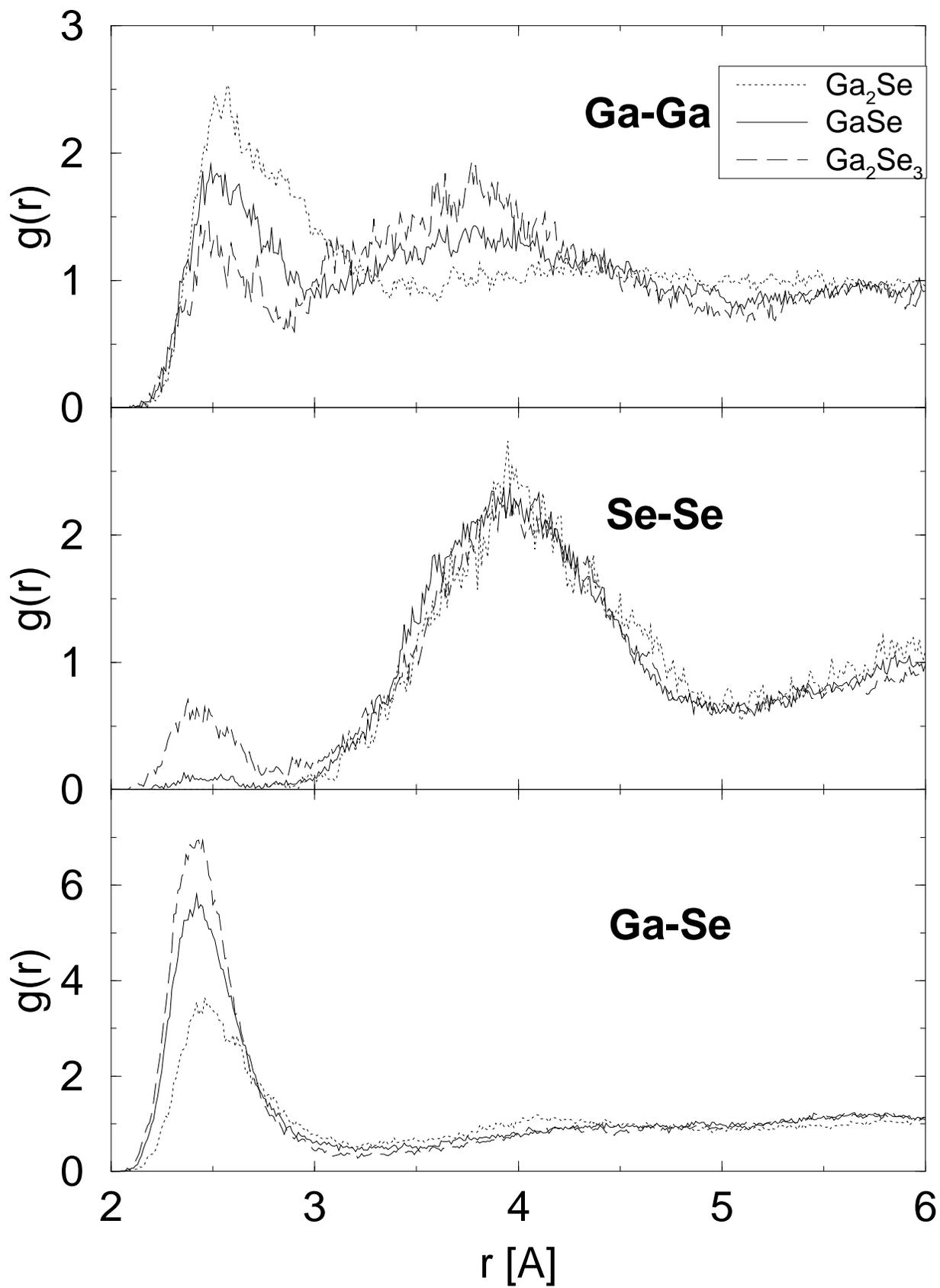
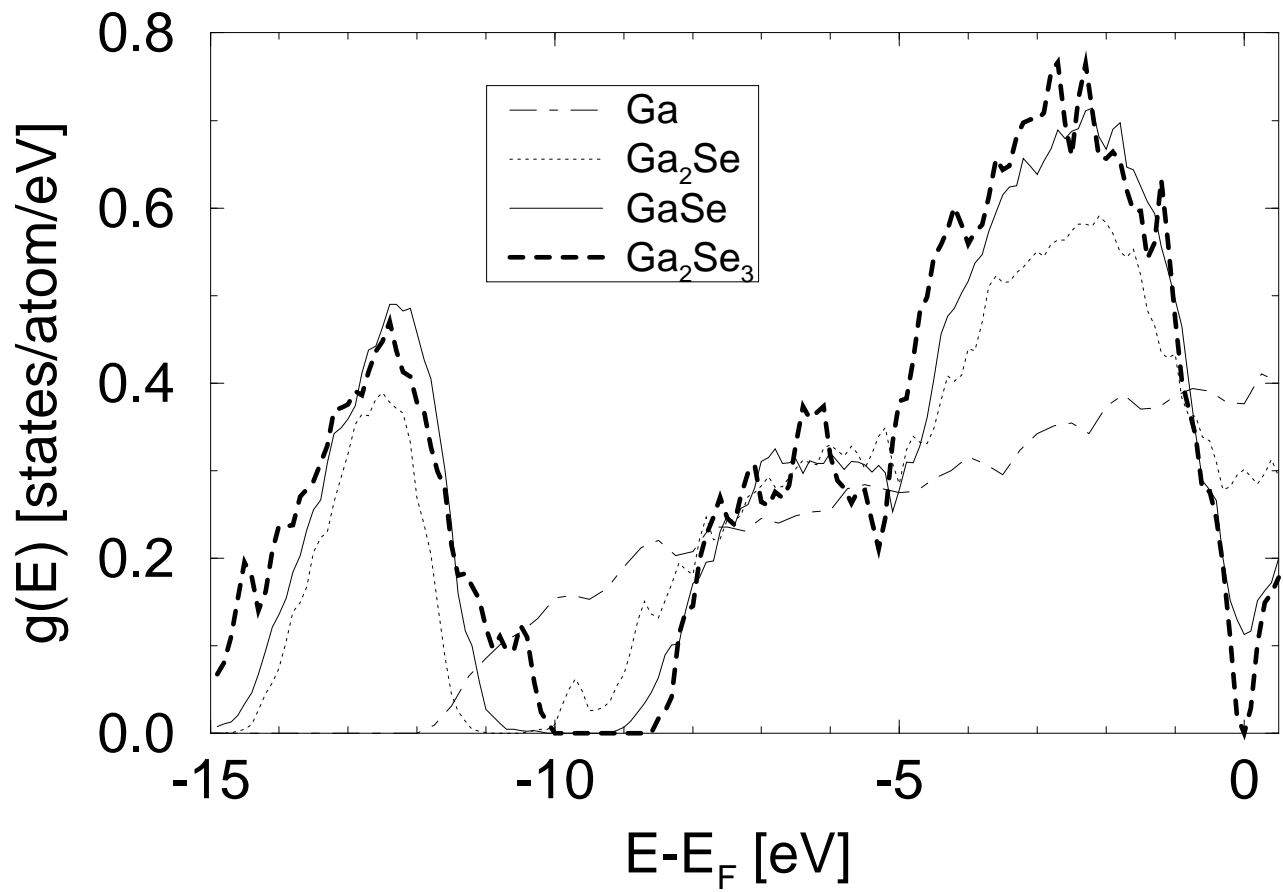
\end{document}